\documentclass{aa}
\usepackage{psfig}

\def\etal {et al.}
\def \lemaitre {Lema\^ \i tre}
\def\eg {{e.g.}}
\def\ie {{i.e.}}
\def\spt {space-time}
\def\Spt {Space-time}
\def\dS{de~Sitter}
\def\rw{Robertson-Walker}
\def\frw{Friedmann-Robertson-Walker}

\def\KKK {Kaluza -- Klein}
\def\frl{Friedmann-Lema\^ \i tre}
\def\eds{Einstein -- de~Sitter}
\def\EdS{Einstein -- de~Sitter}
\def\CUP {Cambridge University Press}
\def\gr {general relativity}
\def\apj#1#2{{Ap.J.}, {#1}, #2}

\def\calH {{\cal H}}

\def\calM {{\cal M}}

\def\Rdot {\dot{R}}

\begin{document}

   \thesaurus{12
             (12.03.2;12.03.4)}
   \title{The Friedmann-\lemaitre ~models in perspective}

   \subtitle{Embeddings of the  Friedmann-\lemaitre ~models in flat 
   5-dimensional space }

   \author{M. Lachi{\`e}ze-Rey
  }

   \offprints{M. Lachi{\`e}ze-Rey}

   \institute{CNRS URA-2052\\ Service d'Astrophysique, C. E. Saclay\\
   91191 Gif sur Yvette cedex, 
France\\      email: marclr@cea.fr
        }

   \date{Received   ; accepted }

   \maketitle

   \begin{abstract}  I show that all FRW models (four dimensional 
pseudo-Riemannian manifolds with maximally symmetric space) can be 
embedded in a flat Minkowski manifold with 5 dimensions. The pseudo 
Riemannian metric of \spt ~ is  induced by  the flat metric.  This 
generalizes the usual embedding widely used for the  \dS ~models. 
I give the coordinate transformations for the embedding. Taking into 
account the spatial isotropy, one can reduce  \spt ~to a 
two-dimensional surface, embedded in a three-dimensional Minkowski 
space. This allows to give exact graphic representations  of the FRW models, and 
in  particular of their curvature.
   \end{abstract}

%

\section{Introduction}

Although immediate experience    indicates that our \spt ~has 
four dimensions,   modern physics evocates additional dimensions in various 
occasions. Gauge theories involve (principal) fiber bundles where the fibers 
may be seen as additional (internal) dimensions where the gauge fields 
live, usually  not considered as physical, since   they do not mix 
with  the \spt ~  dimensions. However, the simplest 
gauge theory, namely the electromagnetism, has been tentatively described 
by a  five dimensional theory  (Kaluza, \cite{Kaluza}; Klein, \cite{Klein}; 
Thiry, \cite{Thiry}). It is not   
clear, in this case,  that  the 5 th dimension may be seen as  a physical one,
but Souriau (\cite{Souriau})  has proposed a   genuine 5 dimensional theory  of 
 this type.

More recently, string theories, M-theory, branes  are formulated in 
a multidimensional \spt. Although   most 
often compactified, the additional dimensions are considered as physical, in the 
sense that    some interactions are  able to propagate through them. 

An appealing property   of the \KKK ~theories is the fact that the 
five-dimensional \spt, in which the Einstein equations are solved, is 
 {\sl Ricci   flat} (and thus devoid of matter), 
although the embedded 4 dimensional manifold ~corresponding to \spt,  our  
world, is curved according to the four-dimensional Einstein equations with 
 sources. 

In this paper, I show that all the Friedmann-\rw ~cosmological 
models can be embedded
in a {\sl flat} (Minkowskian) five-dimensional \spt ~$\calM_{5}$. 
Such an embedding is known for a long time for the de 
Sitter \spt, which appears so as an hyperboloid $\calH$ in $\calM_{5}$. This 
embedding is 
widely used, mainly for  pedagogical and 
illustrative purposes (see, \eg, Hawking and Ellis \cite{Hawking}), and   
presents interesting 
properties for cosmological calculations. Recently, it has for 
instance  been used 
to explore the quantification  on \dS ~\spt ~(Bertola \etal, 
\cite{Bertola}).
Also it is well known that a three dimensional space with maximal 
symmetry can be embedded in a flat Euclidean or Lorentzian manifold, 
also allowing   interesting possibilities for  calculations (see, \eg, 
Triay \etal, \cite{Triay}).

This work can be seen as a generalization of such embeddings  to  \spt 
s with less symmetries (in fact with maximal {\sl spatial} symmetry only). 
All embeddings are in a flat five-dimensional space $\calM _{5} $ with
Lorentzian signature 
(because of this signature, a  flat space does not appear as a 
plane,   as can be seen below). This generalizes also some work made by 
Wesson (\cite{Wesson}) for some peculiar big bang models.
The potential  applications are the same than for the   de Sitter case.
First, this allows to visualize the arbitrary and varying  
curvature of \spt, in the same way  that  \dS ~\spt ~is visualized under the 
form of a hyperboloid embedded in $\calM _{5}$.
I emphasize that   this makes the \spt ~curvature    
visible, not only its  spatial part (which is very simple in all 
cases,  of the three well known types, flat, spherical or 
hyperbolic),   the temporal part being {\sl  not} given by  the curve $R(t)$. 
In Section \ref{geom}, I give explicitly the embedding formulae for 
an arbitrary \spt ~ with maximal spatial curvature, distinguishing   
three cases according to its sign. In  
Section  \ref{dyn}, I consider the cosmic dynamics which, by the 
 Friedmann equations, restricts the geometrical possibilities. I 
 consider in more details some cosmological models. 
 

\section{Cosmology in 5 dimensions}\label{geom}

I recall the metric of   a \frw ~model 
$\calM _{k} $ ,  $$g_{AB}=dt^{2}-R(t)^{2}~d\sigma _{k} 
^{2},~~~~~~~~~~A,B=0,1,2,3$$
where $d\sigma _{k} ^{2}$ is the metric of a maximally symmetric 3-d space 
with curvature  $k=-1,0,1$,$$d\sigma _{k}^{2} = d\chi ^{2}+S_{k} 
^{2}(\chi)~d\omega ^{2},$$ where
$S_{-1}(\chi):=\sinh \chi$,
$S_{0}(\chi):=\chi$,
$S_{1}(\chi):=\sin \chi$. The function $R(t)$ is given by the 
dynamics (see \ref{dyn}). In this section, it will remain arbitrary, so 
that the geometrical embedding appears very general.

I call $\calM_{5}$ the Minkowski \spt ~in five dimensions 
with the  flat metric \begin{equation}
\label{flatmetric}
ds^{2}=\eta _{\mu \nu}~dx^{\mu}~dx^{\nu},~\mu,\nu =0,1,2,3,4.\end{equation}
I will show that every \frl ~model can be seen as a four dimensional 
submanifold (hypersurface) of $\calM_{5}$. 

In the \frl ~models, space has maximal symmetry and is, in particular, 
isotropic. This isotropy is expressed by the action of the group 
SO(3) of  the
spatial rotations $R^{+}( \theta,\phi)$, which coincides with the subgroup 
$G_{(123)}$ of  rotations in the three-dimensional  subspace $\calM_{3}$ of
 $\calM_{5}$ described by 
the coordinates $x^{1},x^{2},x^{3}$.
Thus, any point of $\calM _{k} $ may be written 
$$(x^{0},x^{1},x^{2},x^{3},x^{4})= 
R^{+} (\theta,\phi)~(x^{0},y,0,0,x^{4} ),$$ so that 
$$x^{1}= y ~\cos \phi,$$
$$x^{2}= y ~\sin \phi ~\cos \theta,$$
and $$x^{3}= y ~\sin \phi ~\sin \theta.$$
Thus, in the following, I will simply consider  the three dimensional flat 
manifold  $\calM_{3}$ as embedding space, with 
 the three coordinates $x^{0},x^{1}=y$ and  $x^{4}$ (putting 
 $\phi=\theta=0$), 
 since the others can be trivially reconstructed by action of the 
 spatial rotations. Thus, any point in  $\calM_{3}$ represents a two-sphere 
 in   $\calM_{5}$. 
Space and time are measured in arbitrary identical units (I
   impose $c=1$). In the next sections, I will use 
$H_{0}^{-1}$, the Hubble time, as a common unit. 

\subsection{Negative curvature :}
 I consider $\calM_{-}$ defined parametrically in $\calM_{5}$ through the 
 equations 
 \begin{equation}\label{coordmoins}
x^{0}=R(t)~\cosh \chi 
 \end{equation}
\begin{equation}\nonumber
x^{1}=R(t)~\sinh \chi~\cos \phi
\end{equation}
\begin{equation}\nonumber
x^{2}=R(t)~\sinh \chi~\sin \phi ~\cos \theta
\end{equation}
\begin{equation}\nonumber
x^{3}=R(t)~\sinh \chi~\sin \phi ~\sin \theta
\end{equation}
\begin{equation}\nonumber
x^{4}=\int 
^{t}dt'~\sqrt{\dot{R}^{2}-1}.
\end{equation}
In this case, I have written explicitly the five equations to give 
 insights to the geometry. In $\calM_{3}$, they reduce to 
  $$x^{0}=R(t)~\cosh \chi$$
$$x^{1}:=y=R(t)~\sinh \chi$$
$$x^{4}=\int 
^{t}dt'~\sqrt{\dot{R}^{2}-1},$$ which can be inverted as 
$t=R^{-1}[\sqrt{(x^{0})^{2}-y^{2}}] $ and
$\chi=\sinh  ^{-1} \frac{y}{\sqrt{(x^{0})^{2}-y^{2}}},$
where $R^{-1}$ and $\sinh ^{-1}$ are the inverse functions 
of $R$ and $\sinh $, respectively.

The whole  \spt ~ $\calM_{-}$ is obtained  by the action of the   hyperbolic rotations
 $R^{-}( x^{4},\chi)$
around the $x^{4}$ axis, with angle $  \chi$ (followed by the $SO(3)$ 
spherical rotations $R^{+}(\theta,\phi)$, as indicated above), on the 
world line $\chi=\phi=\theta=0$.
The latter   illustrates the temporal part of the curvature. 
It is defined by its equations 
$$x^{0}=R(t)$$
$$x^{1}=
 x^{2}=  x^{3}=0$$
$$x^{4}=\int ^{t}dt'~\sqrt{\dot{R}^{2}-1}.$$
     
    
\subsubsection{The metric}
 
Differentiation of the   equations (\ref{coordmoins}) leads to 
\begin{equation}
\label{nmetric}
dx^{0}=\dot{R} ~\cosh \chi~dt+R~\sinh  \chi~d \chi
\end{equation}
$$dx^{1}=(\dot{R} ~\sinh  \chi~dt+R~\cosh \chi~d\chi)~\cos \phi
-R(t)~\sinh  \chi~\sin \phi  ~d\phi \nonumber $$
 $$dx^{2}=(\dot{R} ~\sinh  \chi~dt+R~\cosh \chi~d \chi)~\sin \phi~\cos  
 \theta \\
+R(t)~\sinh  \chi~\cos  \phi ~\cos \theta ~d\phi
-R(t)~\sinh  \chi~\sin \phi ~\sin \theta ~d\theta \nonumber $$
$$dx^{3}=(\dot{R} ~\sinh  \chi~dt+R~\cosh \chi~d\chi)~\sin \phi~\sin 
\theta \\
+R(t)~\sinh  \chi~\cos  \phi ~\sin \theta ~d\phi
+R(t)~\sinh  \chi~\sin\phi ~\cos  \theta ~d\theta. \nonumber $$
$$ dx^{4}=\sqrt{\dot{R}^{2}-1}~dt.$$ 

Inserting in (\ref{flatmetric})   leads to the metric induced  onto the surface
$$ds^{2}=   dt^{2}-R(t)^{2}~d\sigma _{-}  ^{2},$$
 \ie, that of a $k=-1$   Friedmann-\lemaitre ~model.
 
\subsection{Positive curvature :}

  I consider $\calM _{+} $ defined parametrically through the equations
\begin{equation}\label{positiveequ}
\label{coordplus}
x^{0}=\int 
^{t}dt'~\sqrt{\dot{R}^{2}+1}
\end{equation}
\begin{equation}\nonumber
x^{1}=R(t)~\sin \chi
\end{equation}
\begin{equation}\nonumber
x^{4}=R(t)~\cos \chi.
\end{equation}
Their inversion leads to 
$t=R^{-1}[\sqrt{(x^{4})^{2}+y^{2}}] $ and
$\chi=\sin ^{-1} \frac{y}{\sqrt{(x^{4})^{2}+y^{2}}}.$

The whole \spt ~  $\calM_{+}$ is obtained by the action of the 
spherical  rotations $R^{+}( 
x^{0},\chi)$,
around the $x^{0}$ axis, with angle $\chi$ [followed by the $SO(3)$ 
spherical rotations $R^{+}(\theta,\phi)$], on the world line 
$\chi=\phi=\theta=0$,
which has the parametric equations 
  $$x^{0}=\int 
^{t}dt'~\sqrt{\dot{R}^{2}+1}$$
$$x^{1}=0$$
$$x^{4}=R(t)   .$$

    
\subsubsection{Metric}

Differentiation  of Eq.(\ref{coordplus}) leads to 
\begin{eqnarray}\label{pmetric}
dx^{0}=\sqrt{\dot{R}^{2}+1} ~dt\\
dx^{1}=(\dot{R} ~\sin \chi~dt+R~\cos \chi~d\chi)\nonumber \\
 dx^{4}=  \dot{R}~\cos \chi~dt-R~\sin \chi ~d\chi\nonumber .
\end{eqnarray}

\subsection{Zero spatial curvature :}
 I consider $\calM _{0} $ defined parametrically through the equations
\begin{equation}\label{coordzero}
x^{0}=[R(t) + \int ^{t}dt'/ \Rdot +R(t)~r^{2}]/2
\end{equation}
\noindent $$x^{1}=R(t)~ r$$
$$x^{4}=[R(t) - \int ^{t}dt'/ \Rdot -R(t)~r^{2}]/2.$$
Their inversion gives
 $t=R^{-1}(x^{0}+x^{4})$ and
$r=\frac{y}{ x^{0}+x^{4}}$.

\subsubsection{Metric}

Differentiation of Eq.(\ref{coordzero}) leads to 
\begin{eqnarray}\label{zmetric}
dx^{0}=
[\Rdot + 1/ \Rdot +\Rdot~r^{2}]~dt/2+R~r~dr \nonumber \\
dx^{1}=\Rdot~ r~dt+R~ dr   \\
dx^{4}=[\Rdot - 1/ \Rdot -\Rdot~r^{2}]~dt/2-R~r~dr.\nonumber 
\end{eqnarray} 

It may be easily  verified that, on the four-dimensional 
hypersurface $\calM_{0}$, this leads   to 
$$ds^{2}=  dt^{2}-R(t)^{2}~d\sigma _{0}  ^{2}.$$

It is advantageous to introduce   the new system of   coordinates:
\begin{equation}
\label{zmetricv} v:=x^{0}+x^{4}=R(t),~x^{1}=R(t)~ r,\end{equation}
  and ~$w :=x^{0}-x^{4}=\int ^{t}dt'/ \Rdot +R(t)~ r^{2}$.\\
 This makes apparent the fact that the whole \spt ~ $\calM_{0}$ 
is obtained by the action of parabolic 
rotations $R^{0}(v,r)$, of angle $r$, 
around the  $v=x^{0}+x^{4}$ axis, of the world-line  
$r=\phi=\theta =0$ (followed by the spatial SO(3) rotations).
The latter  [see an illustration in Fig. (\ref{lambdaSlice})]
is defined 
parametrically  by
$$x^{0}=[R(t) + \int ^{t}dt'/ \Rdot  ]/2$$
$$x^{1}=x^{2}=x^{3}=0$$
$$x^{4}=[R(t) - \int ^{t}dt'/ \Rdot ]/2.$$
The rotation $R^{0}(v,r)$ preserves the value of  the coordinate $v$,   
transforms $x^{1}=0$ to
$x^{1}=R~r=v~r$ and 
$w =  \int ^{t}dt'/ \Rdot$ to $w =  \int ^{t}dt'/ \Rdot + R~r^{2}.$

    
In this representation, (flat) space is represented in $\calM _{3}$ by the 
parabola of parametric equations (\ref{zmetricv}), where $t$ remains 
fixed, or  a paraboloid  in $\calM _{5}$: Fig(\ref{paraboloid}) shows 
this flat space (reduced to two dimensions), 
 in the subspace of $M_{5}$ described by the coordinates   
$x^{1},x^{2},w$. This      (flat) hypersurface  at constant 
time appears as the 
revolution paraboloid obtained by the action of the (spherical) 
rotation around $w$,  of 
angle $\theta$,  of the parabolic section seen above.

It may appear  curious  
that a {\sl flat} space is represented by a parabola 
(or a paraboloid),
rather than by   a straight line (or an hyperplane). This is due 
to the Lorentzian (rather than Euclidean) nature of the embedding 
space $\calM _{3}$ (or $\calM _{5}$). Because of  the signature of the 
metric,     any curve in $\calM _{3}$, with parametric equations  
$x^{0}=f(r),y= A~r, x^{4}=B -f(r)$, 
represents 
a flat space, with  $f$   an arbitrary function, and $A$ and $B$ two 
arbitrary constants. In other words, this curve lies 
in the plane of equation $x^{0}+x^{4}=Ct$, inclined by 45$\deg$ with 
respect to the " vertical " axis. The flat character 
is   expressed by the fact 
that an arc of such a curve corresponding to a range $\Delta y=A~\Delta r$ of 
the coordinate $y$ has precisely $\Delta y$ for length: the contributions due 
to the other coordinates cancel exactly. 
However, for the \frl ~models, the form (\ref{zmetric}) is the 
unique one which gives the complete \rw ~metric. 
   
\subsection{The de Sitter case}
 A peculiar case is the \dS ~\spt, with the topology $S^{3} \times 
 \Re$.     \Spt ~is the hyperboloid 
 $\calH=SO(4,1)/ SO(3,1)$ in $\calM _{5}$ but, as  it is well known,  
 different cosmological models may be adjusted  to it, 
 depending on how the time coordinate is chosen.
 This  gives the opportunity to illustrate the previous cases (all 
 these formulae are standard and may be found, for instance, in 
 Hawking and Ellis, \cite{Hawking}).
 \begin{itemize}
\item  Negative spatial  curvature :  $R(t)=\lambda ^{-1}~\sinh \lambda ~t $.
 
$$x^{0}=\lambda ^{-1}~\sinh  \lambda ~t ~\cosh \chi$$
$$x^{1}=\lambda ^{-1}~\sinh  \lambda ~t ~\sinh \chi$$ 
$$x^{4}=\lambda ^{-1}~\cosh \lambda ~t $$
 
\item Positive spatial  curvature :  $R(t)=\lambda ^{-1}~\cosh \lambda ~t $.
 
$$x^{0}=\lambda ^{-1}~\sinh  \lambda ~t $$
$$x^{1}=\lambda ^{-1}~\cosh \lambda ~t ~\sin \chi$$ 
$$x^{4}=\lambda ^{-1}~\cosh \lambda ~t ~\cos  \chi.$$
   
   \item  Zero spatial  curvature :  $R(t)=\lambda ^{-1}~\exp(\lambda ~t)$.
 
$$x^{0}=\lambda ^{-1}~\sinh  \lambda ~t 
+  \lambda ^{-1}~ (\lambda ~r)^{2}~\exp(\lambda ~t) /2$$
$$x^{1}=\lambda ^{-1}~\exp(\lambda ~t)~ (\lambda ~r)$$ 
$$x^{4}=\lambda ^{-1}~\cosh \lambda ~t 
-  \lambda ^{-1}~ (\lambda ~r)^{2}~\exp(\lambda ~t)/2.$$
Also, for this case,
$$v:=x^{0}+x^{4}=\lambda ^{-1}~\exp(\lambda ~t),$$
$$x^{1}=\lambda ^{-1}~\exp(\lambda ~t)~ (\lambda ~r)$$ 
$$w:=x^{0}-x^{4}= -\lambda ^{-1}~\cosh (- \lambda ~t) 
+  \lambda ^{-1}~ (\lambda ~r)^{2}~\exp(\lambda ~t)/2.$$

Inversion gives
$$t=\lambda ^{-1}~\ln [\lambda ^{-1}~(x^{0}+x^{4})],$$and
$$r=\frac{\lambda ^{-1}~x ^{-1}}{  x^{0}+x^{4} }.$$ 
These coordinates cover half of  the hyperboloid ($x^{0}+x^{4}>0$).
The metric takes the form $ds^{2}=dt^{2} -~\exp(2\lambda ~t)~(d\lambda 
~r)^{2}$,  that of a static universe.

\end{itemize}

Only in  the second case (negative spatial curvature), the \spt ~corresponds to the whole 
hyperboloid, that  I consider  now. 

\subsubsection{Radial light rays}

Light rays are null geodesics with respect to    the metric   of $\calH$. For a 
point $y$ describing a light ray passing through a 
point $x$, we have 
$x~y=(y_{0}-x_{0})^{2}-(y_{0}-x_{1})^{1}-(y_{2}-x_{2})^{2}=0$, and the 
constraints that both $x$ and $y$ belong to $M$.
For a \dS ~universe, this implies  
$$x_{0}^{2}+1=x_{1}^{2}+x_{2}^{2}$$
$$y_{0}^{2}+1=y_{1}^{2}+y_{2}^{2}.$$
After some algebra, this leads to the relations 
$$x_{0}~y_{0}+1=x_{1}~y_{1}+x_{2}~y_{2}$$
and
$$x_{0}~y_{0}+1=x_{1}~y_{2}+x_{2}~y_{1}$$
$$ y_{0}-x_{0}=x_{1}~y_{2}-x_{2}~y_{1}.$$
They describe  a straight line in $\calM _{5}$, which proves that the 
light rays of the (ruled)  hyperboloid are straight lines in $\calM _{5}$.
In  particular,   the light rays through the origin 
$(0,0,1)$  are described by 
$x_{4}=1$ and $x_{0}=\pm x_{1}$.
A similar  treatment shows that,  in the general (non \dS) case,
  the light rays are not, 
in general, straight lines in $\calM _{5}$.

\section{Friedmann equations}\label{dyn}

The  \frw ~universe models obey the Friedmann equation (I use units 
where $c=1$, and $x := R/ R_{0}$)
\begin{equation}
\frac{\dot{x}^{2} }{{x}^{2}~ H_{0}^{2}}=\rho (x) -  \frac{k
 }{R_{0} ^{2}~H_{0} ^{2}~x^{2}  }.
\end{equation} A peculiar model is defined by the 
$R$-dependence of the (dimensionless) density
\begin{equation}
\rho (x)=\Omega _{m}~x^{-3}+\Omega _{r}~x^{-2}+ \lambda~x^{2},
\end{equation}
where   
$\Omega _{m}=8\pi~G~\rho _{0,m}/3H_{0}^{2}$,
$\Omega _{r}=8\pi~G~\rho _{0,r}/3H_{0}^{2}$, 
and $\lambda=\Lambda/3H_{0}^{2}$ are the {\sl present} 
matter density, radiation density and cosmological constant (that I 
include in the density for convenience), in units 
of the critical density $\frac{3~H_{0}^{2} }{8~\pi ~G}$, respectively (additional terms would be 
necessary to represent quintessence).
The dimensionless quantity $\frac{k
 }{R_{0} ^{2}~H_{0} ^{2}}=  \Omega _{m} +\Omega _{r} + \lambda -1$.

\subsection{The spatially flat case}

As an example I   consider the case of the  {\sl spatially} flat  models 
 ($k=0$),   where radiation can be neglected ($\Omega 
_{m}+\lambda =1$). The Friedmann equation takes the simple form
\begin{equation}  \dot{x}^{2} =   H_{0}^{2}~(\Omega 
_{m}~x^{-1} +\lambda ~x^{2}) .\end{equation}
Since $R_{0}$ is arbitrary in this case, I will chose $R_{0}=H_{0}^{-1}$.
I first distinguish two peculiar cases, namely
 \begin{itemize}
\item Empty model with cosmological constant : $\Omega 
_{m}=0$, $\lambda =1$ : the solution is $x=\exp [H_{0}~(t-t_{0})]$.

\item  The \eds ~model, with $\Omega 
_{m}=1$ and  $\lambda =0$. 
The solution is $x= [3~H_{0}~t/2]^{2/3}$, with 
$\dot{x}= (2/3)~[3~H_{0} /2]^{2/3}~t^{-1/3}$,
$H= 2/(3 t)$, and $t_{U}= 2/(3~H_{0})$.
It follows that 
$$\int dt/\dot{R}=  (9/8)~[2/(3~H_{0})]^{2/3}~t^{4/3}.$$

I   show   in Fig.(\ref{paraboloid}) a  spatial  cut of this  \spt.
   The section of \spt ~in the plane $r=0$ is an  inertial world line: 
   this is the   parabola
$w= v^{2}/(2H_{0}^{2})$ shown  by  Fig(\ref{Fig1}).
A perspective of the whole \eds ~\spt ~in M5 is given by Fig(\ref{EdS}).
 
\begin{figure} 
\psfig{file=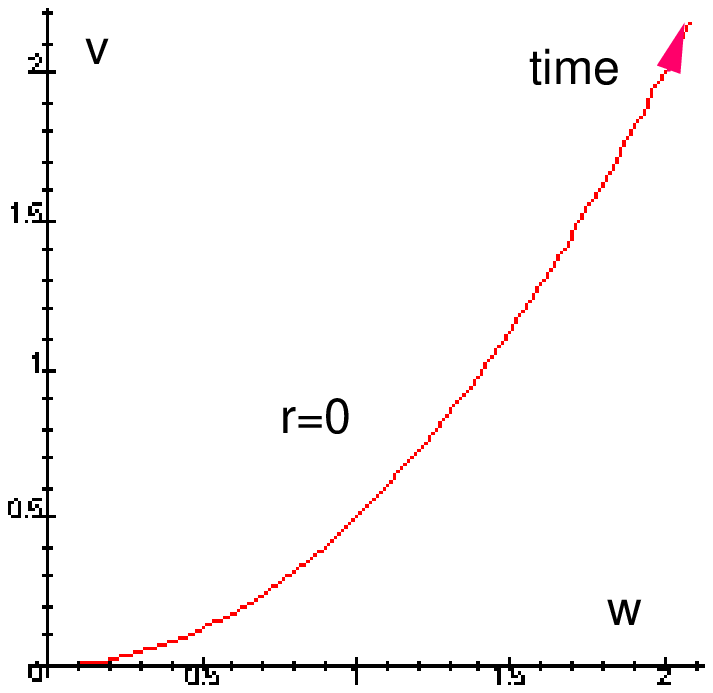,width=8.8cm}
\caption{An inertial world line, \ie, a section $r=0$ of \EdS ~\spt,   
embedded in $M_{5}$}
\label{Fig1}
   \end{figure}

\begin{figure} 
\psfig{file=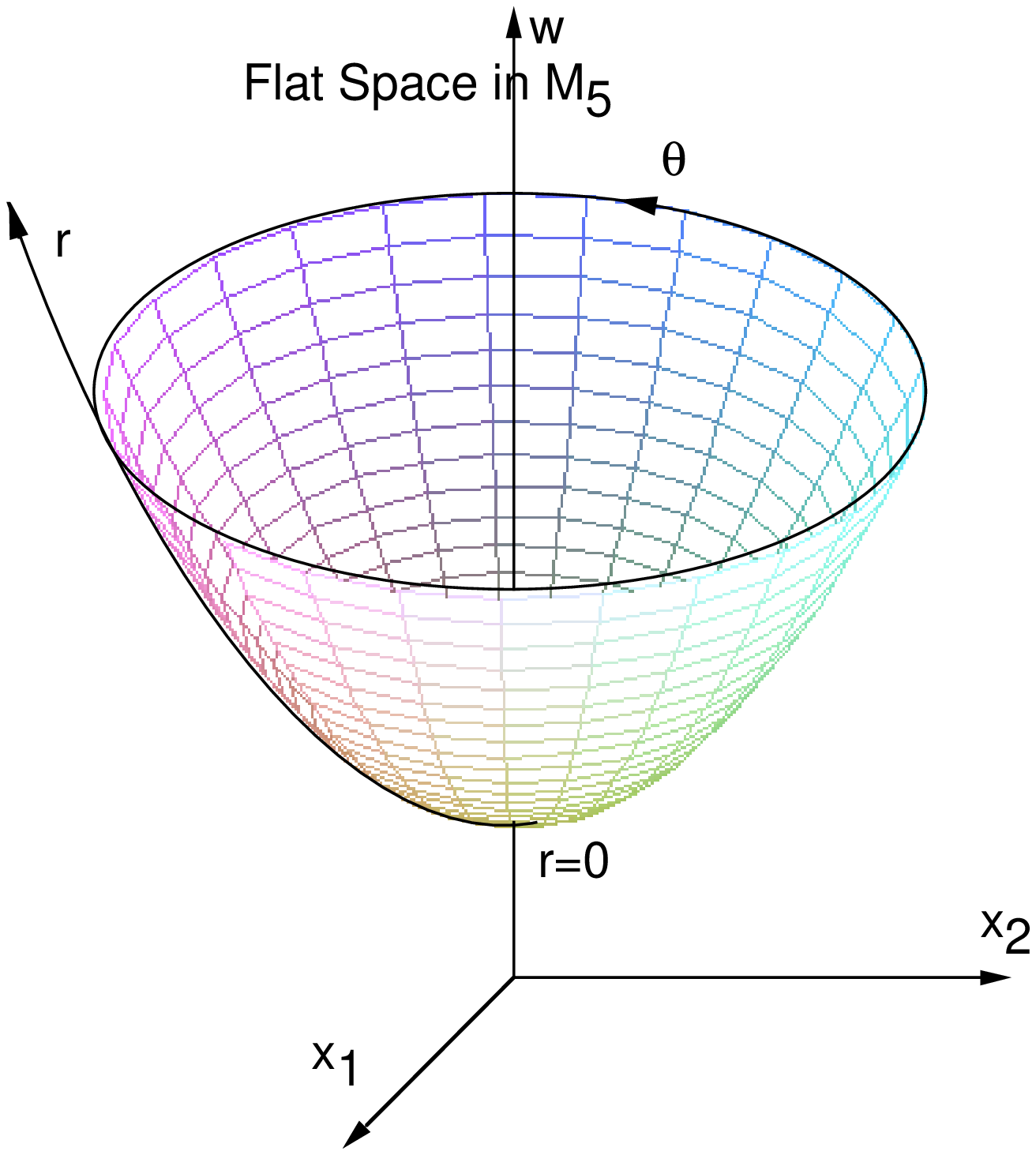,width=8.8cm}
\caption{Flat Euclidean {\sl  space}, embedded in the three-dimensional 
manifold of $M_{5}$ described by the coordinates 
$x^{1}$, $x^{1}$, $w$.}
\label{paraboloid}
   \end{figure}

\begin{figure} 
\psfig{file=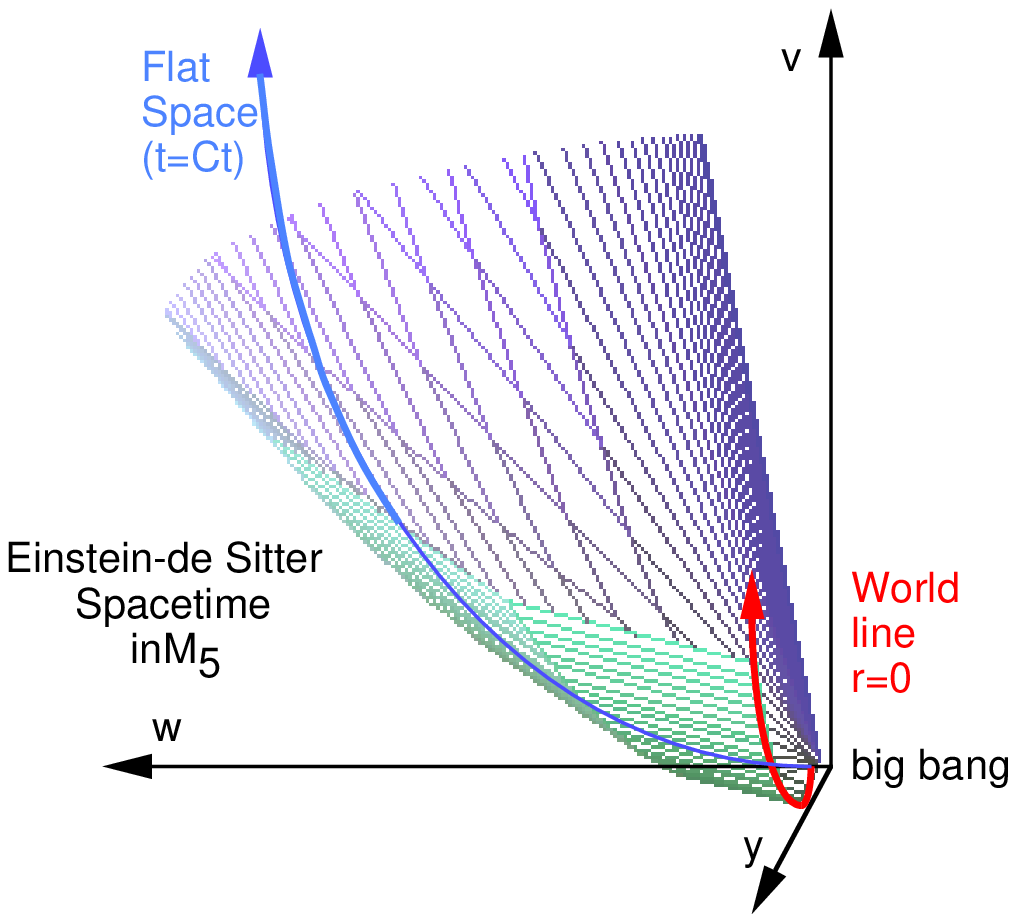,width=8.8cm}
\caption{The Einstein - de Sitter model (with flat   spatial sections) embedded in a flat
 Lorentzian space. Both spatial sections ($t=Ct$) and inertial world 
 lines ($r=Ct$) are parabolas.}
\label{EdS} \end{figure}

\end{itemize}
 
  Now I consider the   general case (spatially flat, assuming $\lambda >0$), 
that I   solve  by  defining $z := \sqrt{ \Omega 
_{m} / \lambda }~x^{3/2}$:  the equation takes the form 
$dz^{2}=\alpha ^{2}~(1+z^{2})~dt^{2}$,
where $\alpha ^{2} := \frac{9~H_{0}^{2}~\lambda}{4}$, with the 
solution $z=\sinh(\alpha~t)$.
Finally, the general solution is 
\begin{equation}
x=(\Omega  _{m} / \lambda ) ^{1/3}~[\sinh (\alpha ~t)]^{2/3}.
\end{equation} All these models have a Big Bang, and I have chosen the 
integration constant so that $x=0$ at $t=0$.
From this, we derive easily
\begin{equation} \dot{x}=(\Omega 
_{m} / \lambda ) ^{1/3}~\frac{2\alpha}{3}~[\sinh (\alpha ~t)]^{-1/3}~
\cosh (\alpha ~t),
\end{equation} and the Hubble parameter $
H(t)=\frac{2\alpha}{3}~\coth (\alpha ~t)$.
 The present period $t_{U}$ corresponds to $x=1$, or $H(t_{U})=H_{0}$, so 
that  $\sinh  \alpha ~t_{U} = \sqrt{\lambda /(1-\lambda)}$.

The $r=0$ section of \spt ~is given by 
$$x^{0}=[R(t) + S(t)]/2$$
$$x^{1}=0$$
$$x^{4}=[R(t) -  S(t) ]/2,$$
where 
$S(t) := \int ^{t}~dt' / \dot{R} (t')$     has  unfortunately no analytical 
expression. 

I recall
$$v:=x^{0}+x^{4}=R(t)$$
$$x^{1}=R(t)~ r$$
$$w:=x^{0}-x^{4}=   S(t) +R(t)~r^{2}.$$
 
\begin{figure} 
\psfig{file=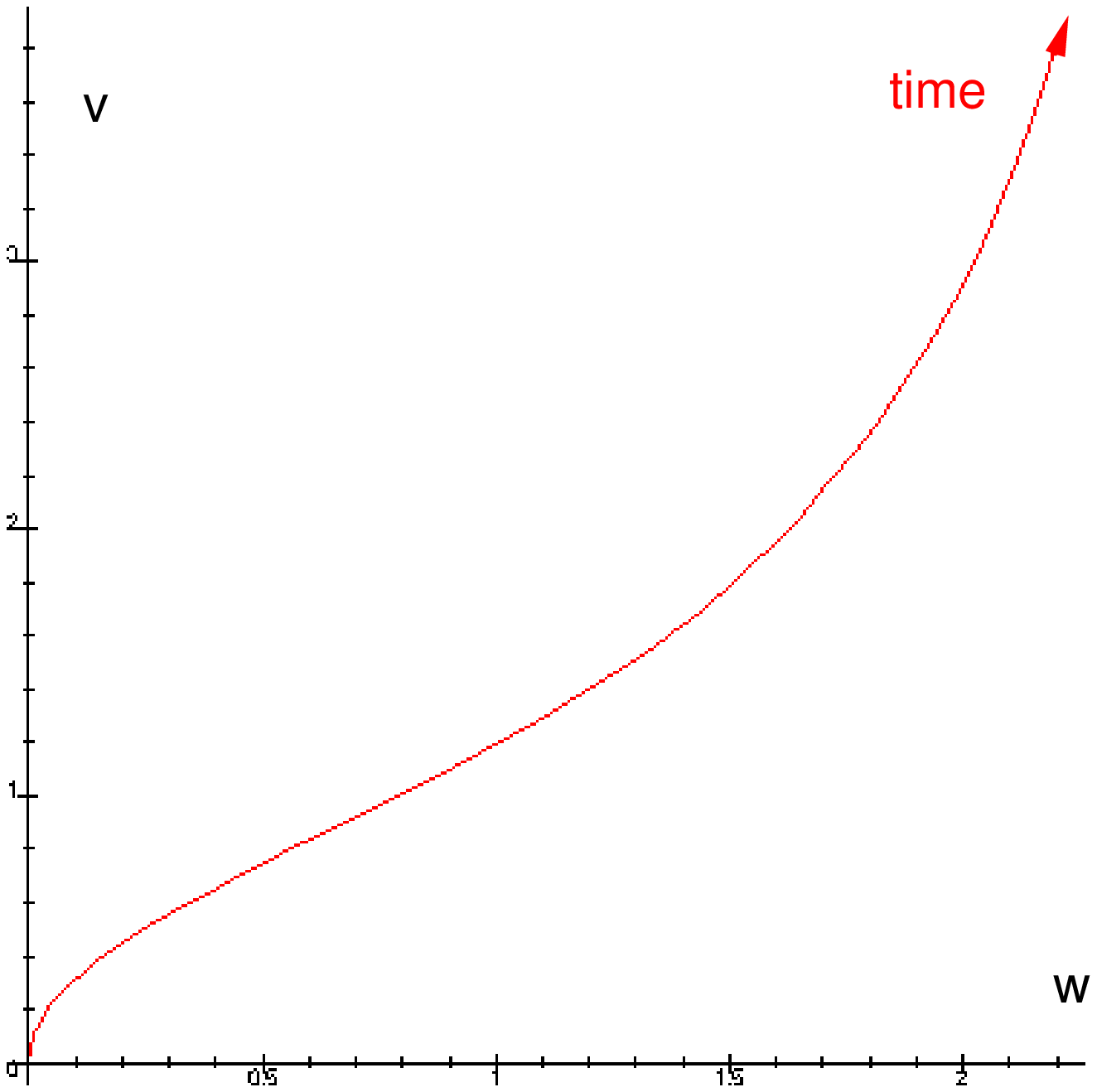,width=8.8cm}
\caption{A world line of the  $\lambda = 2,~ \Omega =2/3$ RW model,   embedded in a flat
 Lorentzian two-dimensional space}
\label{lambdaSlice}
   \end{figure}

\begin{figure} 
\psfig{file=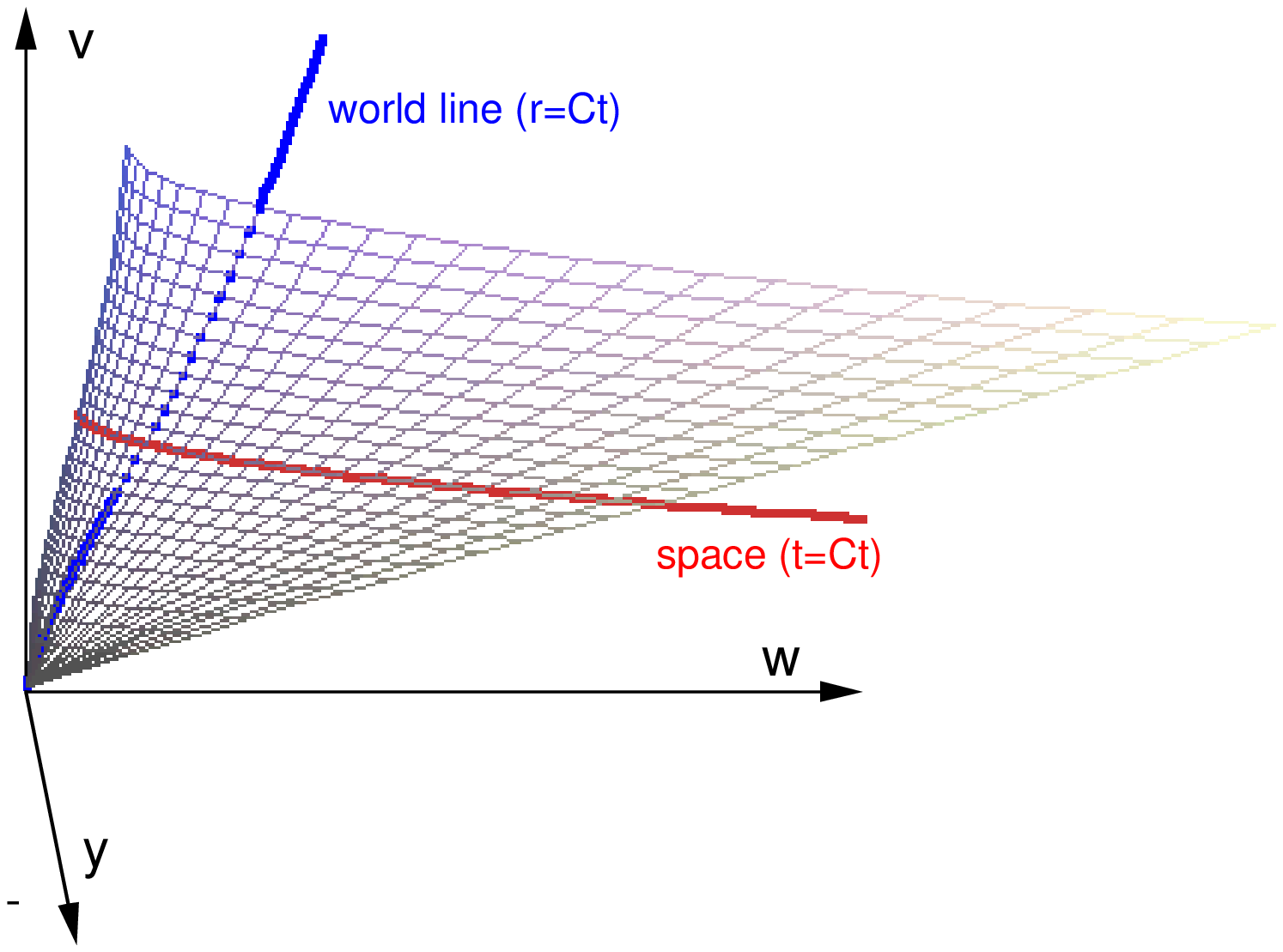,width=8.8cm}
\caption{RW model with $\lambda = 2 \omega =2/3$,   embedded in a flat Lorentzian space}
\label{lambda}
   \end{figure}

 To be more specific, I illustrate (Fig. \ref{lambdaSlice}, \ref{lambda}) the case where $\Omega _{m}=\lambda 
 /2=1/3$, which seems now favored by observational results. Then 
  $\alpha ^{2} := \frac{3~H_{0}^{2}}{2}$, with the solution
  \begin{equation}
x=(1  / 2 ) ^{1/3}~[\sinh (\alpha ~t)]^{2/3},
\end{equation}
\begin{equation} \dot{x}=(1 / 2 ) ^{1/3}~\frac{2\alpha}{3}~[\sinh (\alpha 
~t)]^{-1/3}~
\cosh (\alpha ~t),
\end{equation} and the Hubble parameter\begin{equation}
H(t)=\frac{2\alpha}{3}~\coth (\alpha ~t).
\end{equation}
The present period $t_{U}$ corresponds to
  $\sinh  (\alpha ~t_{U}) = \sqrt{2}$.

\subsubsection{The matter dominated universes}

As an other example, I consider the models where only non-relativistic 
matter governs  the cosmic evolution,  $\rho (x)=\Omega _{m}~x^{-3}$, 
so that  \begin{equation}
\frac{\dot{x}^{2} }{ H_{0}^{2}}=\Omega _{m}~x^{-1} + 1-\Omega _{m}  .  
\end{equation} 

More specifically, I illustrate (Fig.\ref{positive}) a spatially closed model, 
with   $\Omega 
  _{m}=2$. This model (hardly compatible with cosmic observations) 
   has a Big Bang, a maximal expansion at 
  $R=2/H_{0}$ and a Big Crunch. It appears advantageous to use 
  $x^{0}$ as a parameter, with values 0, $R=4/H_{0}$
and $R=8/H_{0}$, respectively, for the three events.
The parametric equations  (\ref{positiveequ})  take the form 
$$  x^{0}$$
$$x^{1}=\sin r~[2-2~(x^{0}/4-1)^{2}],$$
$$x^{4}=\cos r~[2-2~(x^{0}/4-1)^{2}].$$

\begin{figure} 
\psfig{file=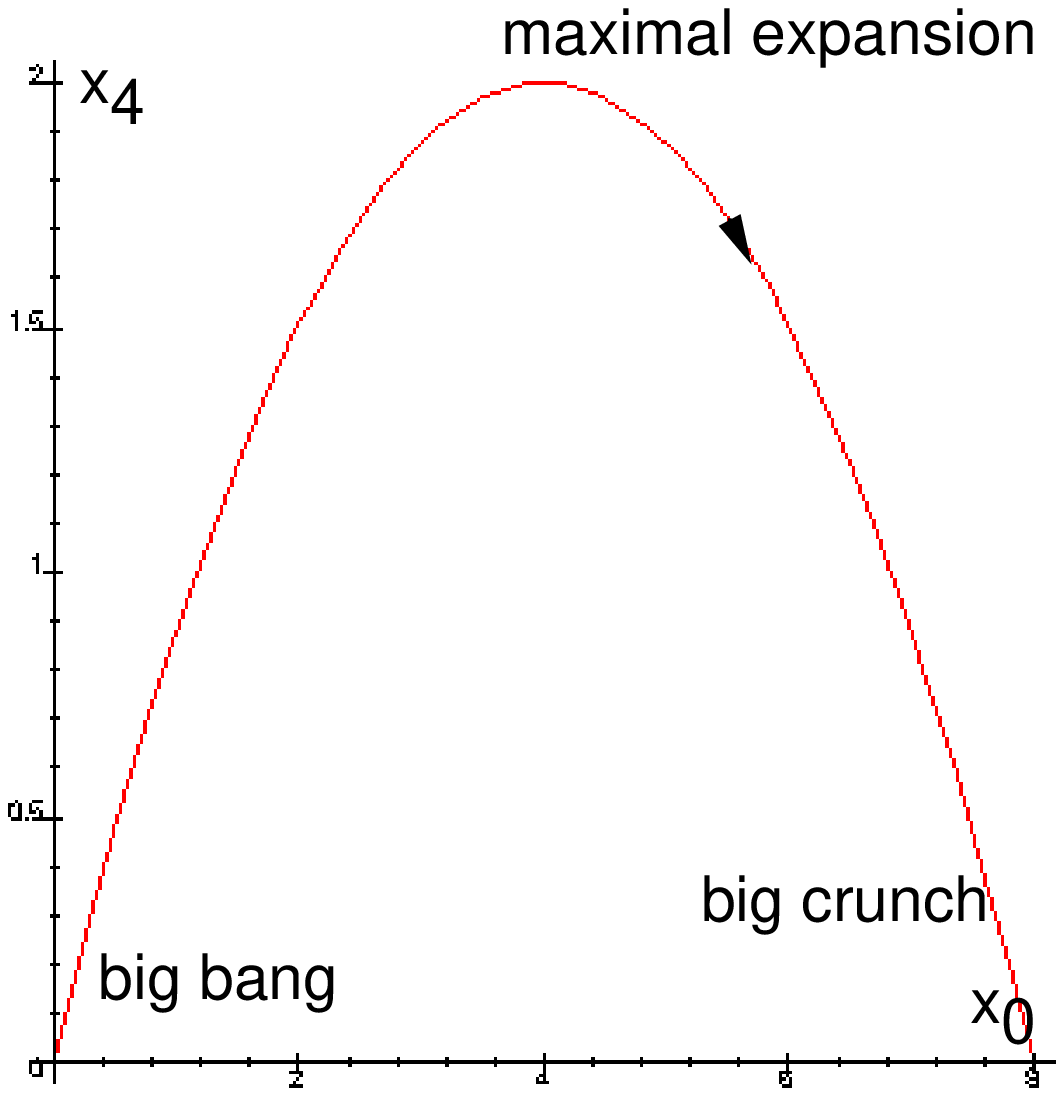,width=8.8cm}
\caption{An inertial world line of a closed RW model, purely matter dominated, with $\Omega 
=2$, is an arc of a parabola.}
\label{PositiveSlice}
   \end{figure}

\begin{figure} 
\psfig{file=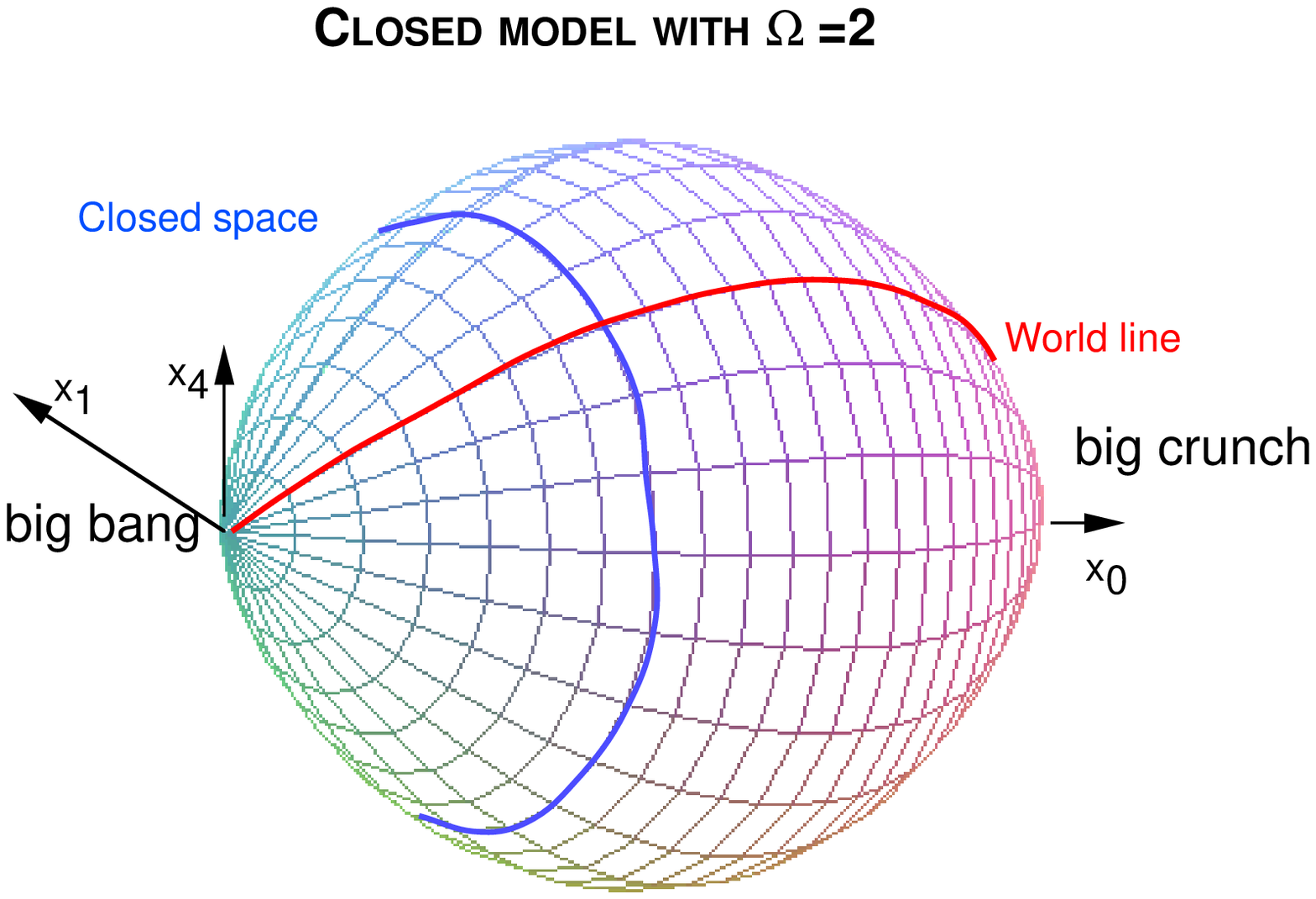,width=8.8cm}
\caption{Closed RW model (purely matter dominated, with $\Omega 
=2$)    embedded in a flat Lorentzian space. Spatial sections ($t=Ct$) 
are circles. Inertial world lines ($r=Ct$) are the arcs of a parabola 
illustrated in Fig(\ref{PositiveSlice}).}
\label{positive}
   \end{figure}

In our representation, the \spt ~is represented by a revolution 
surface of  an arc of parabola [Fig. (\ref{PositiveSlice})]. Spatial sections 
($t=Ct  $) are circles 
(3-spheres in $\calM _{5} $). 
The world lines for inertial particles are the arcs of parabola between 
the Big Bang and the Big Crunch.

\section{Conclusion}

These calculations generalize, to arbitrary \frw ~models, the 
embedding usually used for the \dS ~models. The fact that the embedding
  space is flat offers a very good convenience to illustrate in an 
  intuitive way the geometrical properties of these models. For 
  instance, time durations, or lengths between events  could be 
  obtained by measuring (Lorentzian) lengths of the corresponding  
  curves with a ruler in $\calM _{5} $ or $\calM _{3} $. Also, 
  the curvature coefficients would be  those 
  obtained for the hypersurface in $\calM _{5} $. Care 
  must be taken, however, in this case, that the signature of the 
  embedding space is Lorentzian. Many text books have illustrated  
this fact for the \dS ~case. Here we observe  the curious fact 
that a {\sl flat} space appears as a parabola (or a paraboloid in 
more dimensions). However, a correct measure of the curvature would 
confirm the flatness of the corresponding surface. 

Beside their pedagogical interest, these representations could be of 
great help for various calculations. I mention for instance the 
calculation of cosmic  distances or time intervals 
(generalizing those of  Triay \etal ~(\cite{Triay}) for the case of spatial 
distances). This would be also of great help to gain intuition in any 
theory with more than five dimensions.

Among other speculative ideas, it would be tempting to consider 
dynamics (here cosmic dynamics)   as a geometrical effect in a  
manifold with 5 (or more)  dimensions, which is flat (like here) or 
Ricci flat (this  track is being explored  by 
Wesson, \cite{Wesson}, and references therein). 
  
  This suggest prolongations of the present work in the spirit of the 
  old  \KKK ~attempts: consider other 
  solutions of \gr ~(Wesson and  Liu  \cite{WessonLiu}), consider solutions of 
  gravitation theories other 
  than \gr, consider embeddings in Ricci-flat (rather than flat) 
  manifolds, embeddings in manifolds with more dimensions, etc.  For 
  instance,  Darabi \etal ~(\cite{Darabi}, see also references therein) suggest 
  that this  may  offer a starting point for quantum cosmology.  This may 
also offer an angle of attack for quantization in curved space time, 
following the work already done in \dS ~\spt. This is motivated by recent work
 (see, \eg, Bertola \etal ~2000 and references therein) which have shown 
 interesting relations between quantum field theories in different 
dimensions (for instance, they suggest
the idea that ` a
thermal effect on a curved manifold can be looked at as an Unruh effect
 in a higher 
dimensional flat spacetime `).


\begin{thebibliography}{}

\bibitem[2000]{Bertola}
 Bertola M.,   Gorini V.,   Moschella U.,  Schaeffer R.  2000,
hep-th/ 9906035 v2

\bibitem[2000]{Darabi} Darabi F., Sajko W. N. and 
Wesson  P. S. 2000,  gr-qc/ 0005036
 
\bibitem[1973]{Hawking}Hawking S. W. and Ellis G. F. R., 
{\sl The large scale structure of \spt},  \CUP 1973
 
 \bibitem[1921]{Kaluza} Kaluza Th. Sitz. Preus. Akad., 966, 1921

\bibitem[1926,1927]{Klein} Klein O., Zeits. FŸr Phys., 37, 895, 1926 ;
 Nature, 118, 516, 1927

\bibitem[1963]{Souriau} Souriau J. - M., Nuovo cimento, XXX, 2,  1963

\bibitem[1947]{Thiry} Thiry Y., Journal Math. Pures et Apppl., 9, 1 (1947)

 
\bibitem[1996]{Triay} Triay R., Spinelli L. and Lafaye R. 1996, MNRAS 279, 
564-570

\bibitem[1994]{Wesson} Wesson, P. S.  1994, \apj{436}{547-550}

\bibitem[2000]{WessonLiu}
Wesson P. S. and  Liu H. 2000, gr- qc/ 0003012
 
\end{thebibliography}
 \end{document}